# Critical resolved shear stresses for slip and twinning in Mg-Y-Ca alloys and their effect on the ductility

Mingdi Yu[a], Yuchi Cui[b], Jingya Wang[a,*], Yiwen Chen[a], Zhigang Ding[c], Tao Ying[a], Javier Llorca[d,e,*], Xiaoqin Zeng[a,*]

[a] National Engineering Research Center of Light Alloy Net Forming and State Key Laboratory of Metal Matrix Composites, Shanghai Jiao Tong University, Shanghai 200240, PR China.

[b] School of Materials Science and Engineering, Shanghai Jiao Tong University, Shanghai, 200240, PR China.

[c] Nano and Heterogeneous Materials Center, School of Materials Science and Engineering, Nanjing University of Science and Technology, Nanjing, Jiangsu 210094, China.

[d] IMDEA Materials Institute, 28906 Getafe, Madrid, Spain.

[e] Department of Materials Science, Polytechnic University of Madrid, E. T. S. de Ingenieros de Caminos, 28040 Madrid, Spain.

\* Corresponding author. E-mail address: jingya.wang@sjtu.edu.cn.
\* Corresponding author. E-mail address: javier.llorca@imdea.org.
\* Corresponding author. E-mail address: xqzeng@sjtu.edu.cn.

**Abstract:**

The deformation mechanisms of an extruded Mg-5Y-0.08Ca (wt. %) alloy were analyzed by means of micropillar compression tests on single crystals along different orientations -selected to activate specific deformation modes- as well as slip trace analysis, transmission electron microscopy and transmission Kikuchi diffraction. The polycrystalline alloy presented a remarkable ductility in tension (~32%) and negligible differences in the yield strength between tension and compression. It was found that the presence of Y and Ca in solid solution led to a huge increase in the CRSS for <a> basal slip (29 ± 5 MPa), <c+a> pyramidal slip (203 ± 7 MPa) and tensile twin nucleation (above 148 MPa), while the CRSS for <a> prismatic slip only increases up to 105 ± 4 MPa. The changes in the CRSS for slip and tensile twinning in Mg-Y-Ca alloys expectedly modify the dominant deformation mechanisms in polycrystals. In particular, tensile twinning is replaced by <a> prismatic slip during compressive deformation along the *a-axis*. The reduction of twinning (which generally induces strong anisotropy in the plastic deformation in textured alloys), and the activation of <a> prismatic slip (which provides an additional plastic deformation mechanism with limited hardening) were responsible for the large tensile ductility of the alloy.

**Keywords:** Mg-Y-Ca alloys; micropillar compression; critical resolved shear stress; plastic anisotropy; tension-compression asymmetry; tensile ductility.

## 1. Introduction

Pure Mg and Mg alloys generally present poor ductility and formability, especially at room temperature (Huang et al., 2022; Sun et al., 2019; Tang et al., 2022; Yaghoobi et al., 2022). As a result, forming of rolled sheets and extruded bars becomes difficult and limits the application of wrought Mg alloys in different industrial sectors (Li and Fang, 2022). Thus, understanding the origin of the lack of ductility and formability is of paramount importance to develop new Mg alloys that overcome these limitations.

The poor ductility of Mg alloys is primarily traced to its low-symmetry hexagonal closed packed (HCP) lattice structure, which results in very large differences in the critical resolved shear stress (CRSS) between basal and non-basal slip systems as well as in the easy activation of tensile twinning (Lee et al., 2018). Plastic deformation in pure Mg is initially accommodated by <a> basal slip, which only provides two independent slip systems (Partridge, 1967). This process leads to the development of a strong basal texture during rolling and extrusion. Moreover, plastic deformation along the *c-axis* (which is necessary to activate five independent slip systems to fulfil the von-Mises criterion for homogeneous plastic deformation) is absorbed by tensile twinning, which is triggered at much lower CRSS than that necessary to produce <c+a> pyramidal slip (Graff et al., 2007; Sukedai and Yokoyama, 2010). However, the plastic strain associated with tensile twinning is very limited (at most 7%), moreover, tensile twining is a polar mechanism that only occurs when the stress along the *c-axis* of the crystal is tensile (Mayama et al., 2011). This leads to a large buildup of stresses to activate <c+a> pyramidal slip in grains that are not suitably oriented for twinning and/or that cannot accommodate more plastic deformation by twinning (Obara et al., 1973; Reed-Hill and Robertson, 1957). The stress concentrations in these grains facilitate the nucleation of cracks and limit the ductility (Zhang et al., 2022). Moreover, huge differences in the flow stress and the strain hardening rate between tension and compression appear in textured microstructures, which also lead to fracture during bending and forming operations (Agnew and Duygulu, 2005; Basu et al., 2021).

The strategies to improve ductility and formability of Mg alloys have been directed towards promoting the activation of multiple slip, including non-basal <a> and non-basal <c+a> slip, and to suppress deformation twinning. Multiple slip leads to more homogeneous plastic deformation and limits texture development during rolling and extrusion while twinning promotes plastic anisotropy in textured microstructures (Ahmad et al., 2019; G. Liu et al., 2017; Zhang et al., 2016a). For instance, precipitation hardening in Mg-Zn alloys leads to large enhancements in the CRSS for basal (Alizadeh and LLorca, 2020; Chun and Byrne, 1969; Wang and Stanford, 2015) and pyramidal slip (Alizadeh et al., 2021) and, thus, to an important reduction in the pyramidal-to-basal CRSS ratio. Nevertheless, the large increase in flow stress inherently decreases the ductility due to the strong accumulation of geometrically necessary dislocations around the precipitates (Rosalie et al., 2012). In addition, precipitates also increase the CRSS for twin growth but do not affect the CRSS for twin nucleation (Wang et al., 2019b). As the latter is normally higher than the former, the presence of precipitates do not contribute to hinder the development of twinning. The only difference induced by the precipitates is a larger number of smaller twins, as compared to the precipitate-free condition (Stanford et al., 2012). Thus, precipitate is not very efficient to enhance the ductility of Mg alloys (Fu et al., 2019; Jain et al., 2010).

Strategies based on solid solution hardening have been more successful to improve the ductility of Mg alloys if the alloying elements are properly chosen. For instance, Sandlöbes et al. (Sandlöbes et al., 2013, 2012, 2011) reported that the addition of 3 wt. % Y led to Mg alloys with a tensile ductility > 25 %, which was associated with the presence of a large density of <c+a> pyramidal dislocations in the deformed sample. This behavior was mainly attributed to a reduction in the ratio between the CRSS of the < c+a > pyramidal slip and < a > basal slip, which was ~3.2 according to *in situ* high energy X-ray diffraction tests (Huang et al., 2018; Wang et al., 2018) and ~2.8-4.8 from micropillar compression tests (Wu et al., 2020). Large ductility and formability are not achieved, however, by the addition of other elements in solid solution (such as Al or Zn) because the pyramidal-to-basal CRSS ratio in these alloys are > 10 (Li et al., 2021a; Wang et al., 2020). Zhu et al., (2019) found that the addition of 0.47 wt. % of Ca in solid solution enhanced the activity of <a> prismatic and pyramidal I dislocations as well as the cross-slip between basal and non-basal slip planes, improving the tensile ductility to ~18 % in a Mg-0.47 Ca (wt. %) alloy. And several authors reported a large improvement in the ductility of binary Mg-Zn and Mg-Al alloys through the addition of small amount of Ca (Hofstetter et al., 2015; Sandlöbes et al., 2017; Wang et al., 2021b). This behavior was supported by our recent micropillar compression tests that showed that the addition of Ca to Mg-Zn alloys reduced the pyramidal-to-basal CRSS ratio values, that were similar to those found in Mg-Y alloys (Wang et al., 2021a). Finally, Wu et al., (2018) showed that the presence of Y and Ca reduces the energy for cross-slip/double cross-slip of <c+a> pyramidal dislocations, leading to new dislocation loops which accommodate plastic deformation. In contrast, the cross-slip is inhibited in pure Mg (or in Mg-Al and Mg-Zn alloys) (Wu et al., 2018), by the favorable dissociation of edge pyramidal <c+a> dislocation segments into sessile segments in the basal plane.

Regarding the effect of solid solution on tensile twinning, several investigations reported an increase in the CRSS for twin nucleation and growth with the addition of Al (Wang et al., 2020), Zn (Li et al., 2021a), Y (Li et al., 2021b) as well as Ca to Mg-Zn alloys (Wang et al., 2021a). However, the CRSS for twin nucleation and growth were lower than that for <c+a> pyramidal slip in the corresponding alloy, thus, tensile twinning was still preferred over pyramidal slip to accommodate plastic deformation in grains suitable oriented for twinning. In addition, the addition of 4Y (wt. %) could significantly suppress the tensile twinning (with CRSS larger than 113 MPa) and promote the <c+a> dislocations (with CRSS around 106 MPa) (Wu et al., 2020). The results summarized above point to the beneficial effects of Y and Ca in solid solution to reduce the plastic anisotropy of Mg. Thus, the co-addition of Ca and Y is expected to promote the homogeneous deformation and improve the plastic deformability of Mg alloys, taking advantages of the significant suppression effect of Y on the tensile twinning, the promotion effect of Ca on the non-basal <a> slips, simultaneously the positive effect of Ca and Y on the activation of the <c+a> slips. Ca enhances the activation of <a> prismatic and <a> pyramidal slip while Y has similar effects on <c+a> slip. Moreover, experimental results on the tensile behavior of an extruded Mg – 2.4 wt. % Y – 0.3 wt. % Ca (Zhou et al., 2013) showed a very large tensile elongation (~37 %) but there is not information available in the literature -to the authors' knowledge- on the concurrent effects of Y and Ca in solid solution on the dominant deformation mechanisms and this is the main objective of this investigation. Thus, the CRSS for different slip systems and twinning was determined in a Mg-Y-Ca alloy from micropillar compression tests in single crystals with different orientations. The deformation mechanisms were ascertained from slip trace analysis in the scanning electron microscope (SEM), transmission electron microscopy (TEM) observations of the dislocations as well as transmission Kikuchi diffraction (TKD). This information was used to rationalize the excellent ductility of Mg-Y-Ca and to provide guidelines to design novel Mg alloys with improved ductility and formability.

## 2. Materials and experimental techniques
### 2.1 Materials

The Mg-Y-Ca alloy was prepared from pure Mg (99.99 wt. %), Mg-30 Ca (wt. %) and Mg-30 Y (wt. %) master alloys in a resistance furnace under a protective atmosphere of $CO_2$ and $SF_6$. The actual chemical composition of the ingot, obtained by inductively coupled plasma atomic emission spectroscopy, was Mg-5Y-0.08Ca (wt. %). The cast alloy was solution treated at 400 °C for 12 h, followed by extrusion at 300 °C with an extrusion ratio of ~ 18:1. Afterwards, parallelepipedal samples of 10×10×5 $mm^3$ were cut from the extruded specimens and homogenized at 550 °C for 20 days within quartz capsules filled with Ar to induce grain growth.

**2.2 Experimental techniques**

Tensile and compressive tests were carried out along the extrusion direction in polycrystalline specimens at crosshead speed of 0.5 mm/min, using a universal testing machine (Z100-TEW) at room temperature. The dimensions of the gage section of the dog-bone tensile specimens were 18×3.4×1.4 $mm^3$ (length × width × thickness), while cylindrical specimens of 8 mm in diameter and 12 mm in length were used in the compression tests. Deformation was measured with an extensometer and 3 specimens were tested in each condition.

The crystallographic orientation of the grains in the sample was characterized via electron back-scattered diffraction (EBSD) in a Tescan Mira-3 SEM with an Oxford Instruments Nordlys EBSD detector at an accelerating voltage of 20 kV. The surface of the sample was mechanically ground using abrasive SiC papers with a grit size of 1200, 2000, 3000, 5000 and 7000. Subsequently, the sample surface was electropolished in an ethanol solution with 10 (vol. %) perchloric acid at -30 °C and 30 V for 90 s to remove the surface damage induced by grinding and reveal the grain boundaries. The EBSD data were analyzed using the Channel 5 software and the Oxford Instruments AZtec Nanoanalysis software package v6.0 along with AZtec Crystal. Several grains whose orientations were appropriate to active different deformation modes were selected to mill the micropillars.

Micropillars of 5 × 5 $μm^2$ square cross and an aspect ratio 2:1 were milled from the selected grains using a FEI Helios G4 UX Focused Ion Beam (FIB)/SEM dual beam microscope operated at 30 kV. These dimensions are known to minimize size effects during mechanical deformation while the time and effort to mill each micropillar are reasonable (Wang et al., 2021a). An initial ion current of 9.3 nA was used to remove the surrounding material and it was reduced to 2.5 nA when the beam was getting closer to the actual dimensions of the micropillar. A final ion current of 80 pA was used in the final polishing step to minimize the surface damage due to $Ga^+$ ion-implantation. The final taper of the micropillars was < 1.5°.

Micropillar compression tests were performed in *ex situ* using a Hysitron Triboindenter TI950 system though a diamond flat punch of 10 μm in diameter. All the tests were conducted under displacement control up to a maximum strain of 10 % at a nominal strain rate of $10^{-3}$ $s^{-1}$. The experimental displacement was corrected to account for the elastic deflection of the matrix material beneath the micropillars following the Sneddon correction (Sneddon, 1965). To this end, the elastic modulus of each grain was determined via the nanoindentation method with a Berkovich tip in the same grain where the micropillar was milled. More details about micropillar manufacturing and the compression set-up can be found in (Sneddon, 1965; Wang et al., 2021a).

The engineering stress-strain curves were obtained from the load and the corrected elastic deflection of the micropillar using the initial cross-sectional area and the height of the micropillars measured in the SEM. The yield stress, $\sigma_y$, was determined from the loss of linearity in the stress-strain curve following the methodology described in (Alizadeh and LLorca, 2020; Maaß et al., 2009). From this information, the CRSS of the active slip system was determined as

$$\text{CRSS} = \text{SF} \times \sigma_y \tag{1}$$

where SF is the Schmid factor of the corresponding slip system, computed from the crystallographic orientation of each crystal (Table 1).

The slip traces on the top and lateral surfaces of the deformed micropillars were characterized in a Tescan Mira-3 SEM to ascertain the active slip planes. The active slip plane and direction were identified from the micropillar orientation using VESTA software (Momma and Izumi, 2008). Moreover, TEM and TKD were used to determine the dislocation activity and the orientation of the micropillar after deformation. To this end, a thin lamella was lifted-out along the loading direction from the deformed pillars and thinned to < 100 nm in thickness using FIB. The TKD maps were collected in a Tescan Mira-3 SEM at 30 kV with a step size of 20 nm. The TEM observations were carried out using a Talos F200X G2 microscope operated at 200 kV. The two-beam condition was applied to obtain dislocation contrast. Moreover, the "g·b" visibility criterion was used to identify the types of dislocation, i.e., the dislocation is in contrast when $\vec{g} \cdot \vec{b} \neq 0$, where $\vec{g}$ is the diffraction vector and $\vec{b}$ the Burgers vector.

**2.3 First-principles calculations**

In order to study the influence of Y and/or Ca atoms on the deformation mechanisms in Mg alloys, the generalized stacking fault energy (GSFE) curves of different slip systems were calculated via the first-principles calculations using the Vienna Ab initio Simulation Package (VASP) (Kresse and Furthmüller, 1996). The exchange-correlation function was described using the generalized gradient approximation (GGA) with the Perdew-Burke-Ernzerholf functional (PBE), based on the projector augmented wave (PAW) (Blöchl, 1994) method.

A supercell with 12-layers containing 48 atoms was defined for different slip systems, as indicated in Fig. 1. Each supercell was separated by 15 Å vacuum to eliminate the influence of the periodic boundary conditions. The formation energy was initially calculated for different positions of the solute atoms and the configurations with lower formation energy was selected as the most stable ones (Yuasa et al., 2014). In the binary $Mg_{47}N_1$ (N = Y, Ca) alloys, the most stable configuration was found when one Mg atom at the center site of the stacking fault plane was substituted by a solute atom X. In the ternary $Mg_{46}N_1X_1$ (N = Y, and X = Ca) alloy, the most stable configuration was found when one Mg atom at the center site of the stacking fault plane was substituted by a Ca atom. Then, one of the eleven nearest Mg atoms from the Ca atom was substituted by one Y atom, as shown in Fig. S1 in the supplementary material. The exact position of the Y atom was determined from the formation energy (Ding et al., 2019; Dong et al., 2018). The formation energies for every occupancy of the Y atom are listed in Table S1 in the supplementary material.

The conventional direct crystal slip methods were employed to obtain the GSFE curves of different slip systems The perfect supercell was cut into two free parts and one part was displaced with respect to the other one along the slip direction. The atomic positions were relaxed only along the direction perpendicular to the stacking fault plane (Wang et al., 2020). A residual force threshold of 0.01 eV/Å was performed in all geometric relaxations until the electronic energy converged to less than $10^{-5}$ eV/cell. The Brillouin zone for the GSFE of the basal slip system, the prismatic slip system, and the pyramidal slip system was set as 8×8×1, 10×6×1, and 6×10×1, respectively, with an energy cutoff of 480 eV (Dong et al., 2018; Wang et al., 2013).

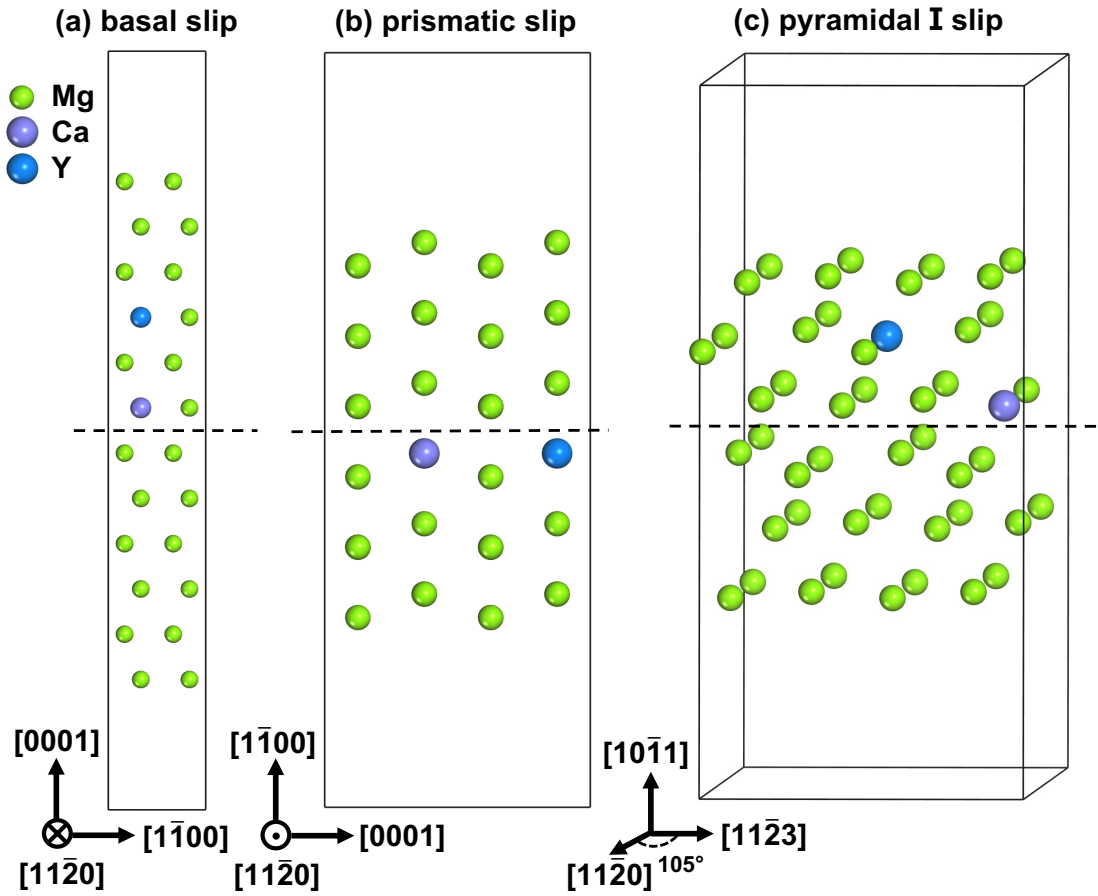

**Fig. 1.** Schematic illustration of the models to calculate the GSFE for (a) basal slip (b) prismatic slip, and (c) pyramidal I slip. The most stable positions of Y and Ca atoms determined by the lowest formation energy are marked by blue and purple atoms, respectively. Stacking fault planes are noted by the dotted lines.

## 3. Results

### 3.1 Mechanical behavior of polycrystals

The inverse pole figure (IPF) map of the as-extruded Mg-Y-Ca alloy along the extrusion direction is plotted in Fig. 2a. The {0001} pole figure shows that the Mg-Y-Ca alloy possesses a weak texture with a strength of ~ 8.21 mrd, as displayed in Fig. 2b, compared to pure wrought Mg with a strong basal texture of >15 mrd (Yin et al., 2021). The engineering stress-strain curves of the extruded Mg-Y-Ca alloy from the tensile and compressive tests parallel to the extrusion direction are plotted in Fig. 2c. The scatter was very limited and the average tensile elongation was very large (≈ 32%). Moreover, the tensile yield stress was 104 MPa, very close to the yield strength in the compression tests (122 MPa). Thus, the Mg-Y-Ca alloy presented very low tension/compression asymmetry in the yield strength in contrast with the marked asymmetry in extruded Mg and Mg alloys (Sukedai and Yokoyama, 2010; Yin et al., 2021; Zhang et al., 2016b).[1] It should also be noted that volume fraction of the twinned material after tensile deformation was very low (≈ 1.8%), indicating that twining was not a dominant deformation mechanism in the Mg alloy.

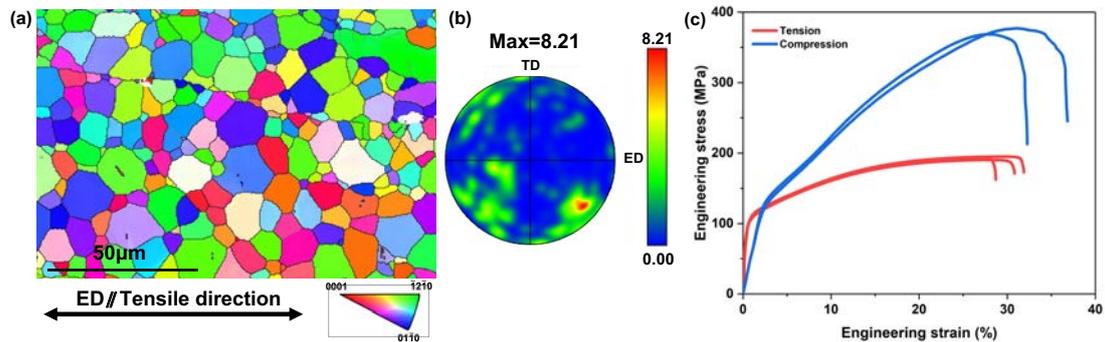

**Fig. 2.** (a) IPF map of the Mg-Y-Ca along the extrusion direction. (b) {0001} Pole figure of the Mg-Y-Ca alloy illustrating the texture characteristics before the deformation in the TD-ED plane. (c) Engineering stress-strain curves in tension and compression parallel to the extrusion direction of the Mg-Y-Ca alloy.

### 3.2 Deformation mechanisms

---

[1] The comparison between both curves shows the limited tension-compression anisotropy in the yield strength but the differences in the elastic and fully plastic regions are due to the limitations of the compression tests. Compression tests always underestimate the elastic modulus because it is very difficult to ensure that the specimen surface and the loading plate surface are perfectly parallel. Thus, partial contact between both surface leads to localized plastic deformation and to an apparent elastic modulus that is lower than the real one. Moreover, barreling of the cylindrical specimen during compression leads to non-homogeneous plastic deformation and overestimates the strain hardening for large plastic strains.

The IPF map with the crystallographic orientation of the grains in the Mg-Y-Ca alloy is depicted in Fig. 3. The grains were larger than 150 μm, and the micropillars were milled from the center of the grains to ensure that they were single crystals. Four grains with appropriate orientations (Fig. 3) were selected to activate different deformation mechanisms. The loading directions in the four grains are listed in Table 1, as well as the maximum Schmid Factor (SF) for the corresponding slip systems (<a> basal slip, <a> prismatic slip, <a> pyramidal I slip, <c+a> pyramidal I slip and <c+a> pyramidal II slip) as well as $\{10\bar{1}2\}$ tensile twinning. The inclination angle in Table 1 indicates the angle between the *c-axis* of each grain and the compression direction, as presented. The compression direction is nearly parallel to $[11\bar{2}0]$, $[10\bar{1}0]$, and $[0001]$ in grains B, C and D, respectively, and forms an angle of ~ 48.5° with respect to $[0001]$ axis in grain A. Herein, grain A presents the highest SF for <a> basal slip, which is prone to be the dominant deformation mechanism during compression. Plastic deformation along the <c+a> pyramidal I and II systems is favored in Grain D. <a> prismatic and pyramidal as well as <c+a> pyramidal slip systems have similar SFs in grain B, while grain C is suitably oriented to promote tensile twinning and <a> prismatic slip.

Table 1. The loading direction, inclination angle, elastic modulus, and maximum Schmid factor for each slip system and tensile twinning in the selected grains.

| Grain | Loading direction | Inclination angle (°) | Elastic modulus (GPa) | Maximum Schmid factor ||||||
|---|---|---|---|---|---|---|---|---|---|
| | | | | Basal <a> | Prismatic <a> | Pyramidal I <a> | Pyramidal I <c+a> | Pyramidal II <c+a> | Tensile twin |
| A | $[11\bar{2}3]$ | 48.5 | 46.04 | 0.44 | 0.25 | 0.42 | 0.36 | 0.20 | 0.17 |
| B | $[11\bar{2}0]$ | 83.4 | 48.14 | 0.11 | 0.48 | 0.46 | 0.47 | 0.47 | 0.43 |
| C | $[10\bar{1}0]$ | 87.8 | 46.48 | 0.03 | 0.46 | 0.42 | 0.43 | 0.37 | 0.49 |
| D | $[0001]$ | 4.5 | 47.47 | 0.06 | 0.00 | 0.03 | 0.44 | 0.47 | -* |

*Tensile twinning cannot be activated during compression along the *c-axis*.

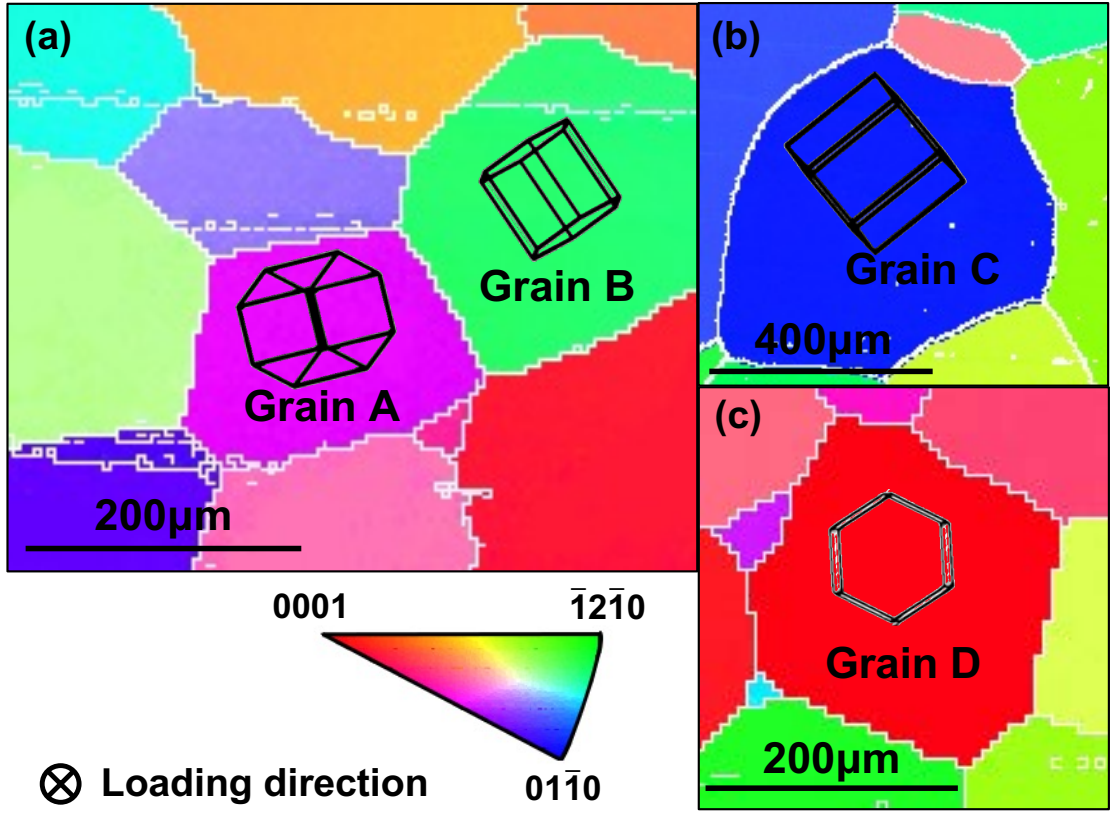

**Fig. 3.** Inverse pole figure (IPF) map showing the crystallographic orientation of the grains in the Mg-Y-Ca alloy. (a) The loading direction in micropillars from grains A and B form an angle of ~ 48.5° with the [0001] crystal orientation and are parallel to [11$\bar{2}$0], respectively. (b) The loading direction in micropillars from grains C is parallel to [10$\bar{1}$0]. (c) The loading direction in micropillars from grain D is parallel to [0001]. The compression loading direction is perpendicular to the paper.

### 3.2.1 Deformation mechanisms in micropillar of grain A

The engineering stress-strain curves obtained from the compression micropillars carved from grain A along [11$\bar{2}$3] orientation are plotted in Fig. 4a. For the sake of clarity, the horizontal axis of the green curve in Fig. 4a is shifted by 0.5%. After the initial elastic region, the curves show gradual yielding and reach a plateau in the flow stress at an applied strain of ~ 5%, without significant work hardening afterwards. This behavior is consistent with a plastic deformation dominated by basal slip in pure Mg and Mg alloys (Kiener et al., 2021; Y. Liu et al., 2017; Luo et al., 2022; Wang et al., 2020, 2019a; Wu et al., 2020). Small strain bursts (noticed by sudden drops in the stress) are present in the stress-strain curves and they are associated with the activation of dislocation sources in particular basal slip planes. However, the magnitude of the strain bursts is much smaller than that reported in other Mg alloys. In fact, large strain bursts are associated with the localization of deformation in a few slip planes along the micropillar (Wang et al., 2019). However, the lateral and top views of the micropillar after deformation (Figs. 4c and 4d, respectively) show evidence of uniform slip traces along the length and width of the micropillar, indicating that plastic deformation was homogeneous. A yield stress of 65 ± 11 MPa (indicated by the black stars in the inset of Fig. 4a) was determined from the critical points in the engineering stress-strain curves when the curves deviated from linearity, following the procedure detailed in (Alizadeh and LLorca, 2020; Wang et al., 2019a).

Secondary electron images of lateral and top views of the deformed micropillars were obtained in the SEM to ascertain the actual deformation mechanisms and are shown in Figs. 4c and 4d, respectively. Many parallel slip traces appear on the top and lateral surfaces, which were not present before deformation (Fig. 4b). The orientation of the slip traces on the micropillar surfaces is indicated by the green dashed lines in Figs. 4c and 4d. The slip steps were obviously observed on the top view from the top right corner to the lower left corner, and the corresponding slip direction is determined as marked with a white arrow in Fig. 4d. They are indicated by blue planes and red arrows, respectively, in Figs. 4e and 4f within the crystallographic lattice. It is evident that the slip traces in the micropillar are parallel to the basal planes and the shear deformation takes place along the $[2\bar{1}\bar{1}0]$ direction, as shown from the top and lateral views of the deformed micropillar. In fact, the (0001) <$2\bar{1}\bar{1}0$> basal slip system has highest SF (listed in Table 1) and plastic deformation along this slip system is dominant in this micropillar. Therefore, the CRSS for <a> basal slip (based on the yield stress and the corresponding SF) can be estimated as 29 ± 5 MPa.

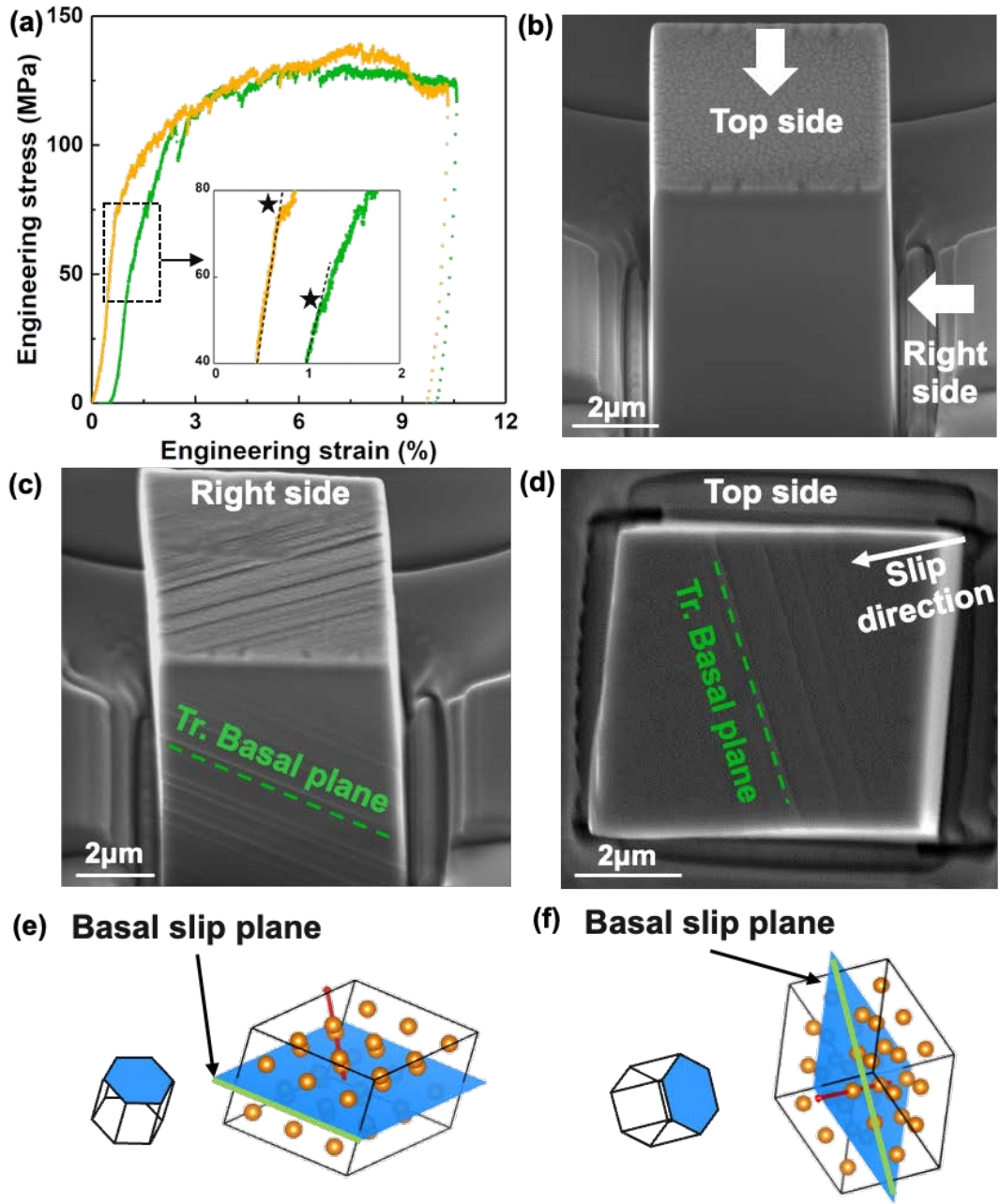

**Fig. 4.** (a) Engineering stress-strain curves obtained from micropillar compression tests in grain A. The yield stress is marked with a black star. SEM images of the micropillar (b) before compression and after compression from the (c) right lateral view and (d) top view. The slip plane trace and slip direction are indicated by the green dashed lines and the white arrow, respectively. The schematic crystallographic lattice of the corresponding slip plane is presented (e) for right side and (f) for top side. The blue planes indicate the theoretical basal glide planes, and the red arrows represent the corresponding shear directions.

### 3.2.2 Deformation mechanisms in micropillars of grains B and C

Representative engineering stress-strain curves obtained from micropillar compression tests along [11$\bar{2}$0] in grain B and along [10$\bar{1}$0] in grain C are plotted in Figs. 5a and 5b, respectively. The horizontal axis of the green and blue curves was shifted by -0.1% and +0.1%, respectively, in the inset of Fig. 5b for the sake of clarity. The stress-strain curves are smooth, without distinct strain bursts. The initial elastic region is followed by another linear plastic region with reduced strain hardening rate. This behavior is radically different from that observed in micropillars with equivalent orientation in pure Mg and several Mg alloys (Mg-Al, Mg-Zn, Mg-Y and Mg-Zn-Ca) (Li et al., 2021b, 2021a; Y. Liu et al., 2017; Wang et al. 2021a, 2020; Wu et al., 2020), which presented large strain bursts after the initial elastic region due to the nucleation of tensile twins at the top of the micropillar. They are similar to those found in Mg-2Y (wt. %) alloy at 250 °C (Li et al., 2021b), where <a> prismatic slip replaced twinning as the dominant plastic deformation mechanisms. The yield stresses (obtained as indicated above and marked with purple stars in Fig. 5) were 219 ± 9 MPa and 228 ± 4 MPa along [11$\bar{2}$0] and [10$\bar{1}$0] orientations, respectively.

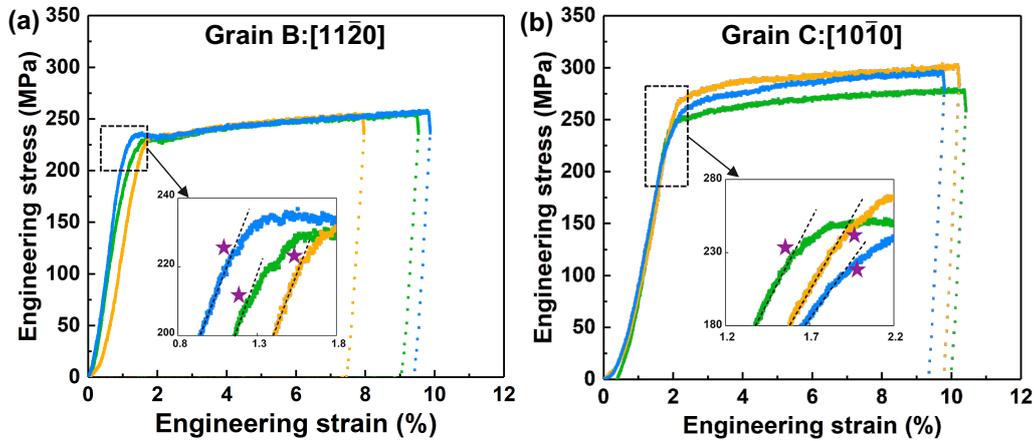

**Fig. 5.** (a) Engineering stress-strain curves obtained from micropillar compression tests in grain B along [11$\bar{2}$0]. (b) Idem in grain C along [10$\bar{1}$0].

The representative morphology of the micropillar deformed along [11$\bar{2}$0] (grain B) is depicted in the SEM images in Figs. 6a and 6b from two different sides (front and left, respectively). Faint slip traces are visible on both lateral surfaces of the deformed micropillars, as indicated by the blue dashed lines in Figs. 6c and 6d, which show the rectangular zones marked by dashed lines in Figs. 6a and 6b, respectively, at higher magnification. The slip traces are distributed homogeneously along the lateral surfaces, indicating that plastic deformation was uniform along the micropillar. Moreover, there are not slip steps at the surface (as opposed to the micropillar deformed along [11$\bar{2}$3] in Figs. 4c and 4d), in agreement with the smooth stress-strain curves. This deformation morphology is different from that observed in other Mg alloys compressed along *a-axis* (Li et al., 2021b, 2021a; Y. Liu et al., 2017; Wang et al., 2020; Wu et al., 2020), where two regions with different contrast were always observed after the deformation due to the nucleation of tensile twins.

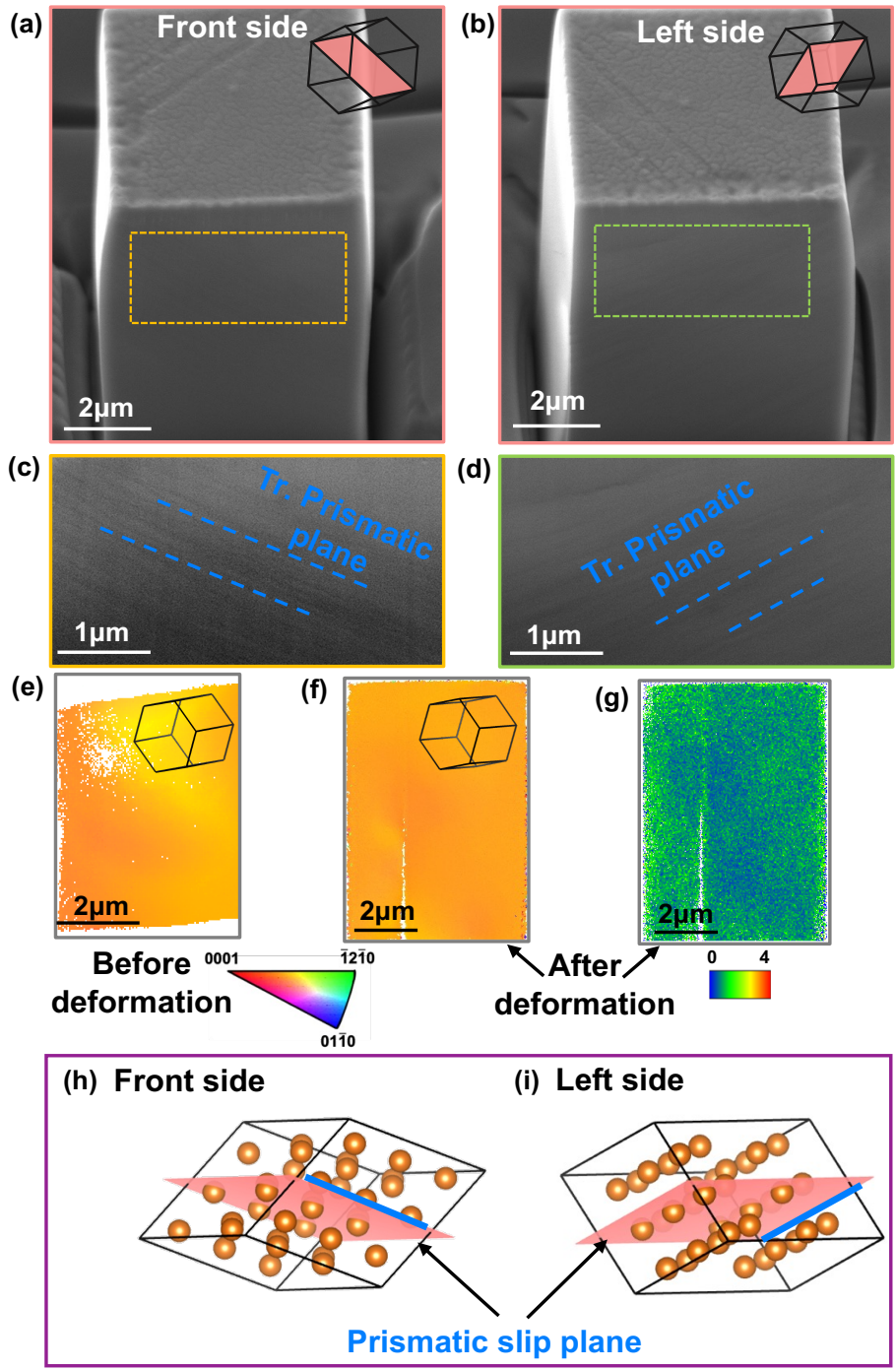

**Fig. 6.** SEM images of the micropillar deformed along [11$\bar{2}$0] from grain B. (a) Lateral front and (b) lateral left view side. The traces of the active slip planes are indicated by the blue dashed lines in Figs. 5c and 5d, which show the rectangular zones marked by dashed lines in Figs. 5a and 5b, respectively, at higher magnification. (e) and (f) TKD maps of the lamella extracted from the undeformed region in grain B and along the compression direction from the deformed micropillar, respectively. (g) KAM map of the deformed micropillar. (h) and (i) Schematics of the crystallographic lattice showing the corresponding slip plane for the lateral front side and left side, respectively. The red planes indicate the theoretical prismatic glide planes, and the blue lines represent the corresponding slip traces.

In order to identify the deformation mechanisms, two parallel thin foils were extracted from the undeformed region in grain B and along the loading direction from the deformed micropillar, respectively, and their orientation was determined by TKD. The position of the lamellae is indicated in Fig. S2 of the supplementary material. The corresponding orientation maps in Figs. 6e and 6f show that the IPF map (∥Z) of the undeformed and deformed thin foils share the same orientation and demonstrate that tensile twins were not nucleated during micropillar compression up to 10% strain. Moreover, Fig. 6g presents the kernel average misorientation (KAM) map of the whole pillar in Fig. 6g reveals the homogeneous deformation without shear bands assuming an angular threshold of 4°. The slip traces on the lateral surfaces of the micropillars were associated with the prismatic planes, as indicated in Figs. 6h and 6i. Thus, prismatic slip was triggered at the onset of the yielding and dominated plastic deformation. The maximum SFs for <a> prismatic slip, <a> pyramidal I, <c+a> pyramidal II slip and tensile twinning were very similar along this orientation (Table 1) but the presence of Y and Ca in solid solution favored the activation of prismatic slip. It should be noted that <c+a> pyramidal I slip dominated plastic deformation and hindered the development of tensile twinning in micropillar compression tests along [$\bar{1}$2$\bar{1}$0] orientation in a Mg-4Y (wt. %) (Wu et al., 2020). The maximum SFs for <c+a> pyramidal I slip, tensile twinning and <a> prismatic slip in this orientation were 0.41, 0.46 and 0.49 and, thus, the preference of <c+a> pyramidal slip can be associated with the higher CRSSs for tensile twin nucleation and <a> prismatic slip in Mg-4Y alloy (Wu et al., 2020).

Similar deformation morphology was found in the micropillars deformed along [10$\bar{1}$0] in grain C. Continuous slip traces were homogeneously distributed along the lateral surfaces, as indicated by the dashed blue lines from in Fig. 7b, which shows the rectangular region marked by dashed lines in Fig. 7a at higher magnification. As in the previous case, the micropillar orientation before and after deformation was assessed by TKD carried out in a thin lamella extracted from the undeformed region (Fig. 7c) and from the deformed micropillar along the loading direction (Fig. 7d), respectively. The relative orientation between the two thin lamellas is shown in Fig. S3 in the supplementary material. The corresponding orientation maps do not show any evidence of tensile twinning and <a> prismatic slip was again the dominant plastic deformation mechanism. This conclusion is supported by the uniform plastic deformation without obvious shear bands revealed by the KAM map assuming an angular threshold of 2° (Fig. 7e) and the agreement between the slip traces on the lateral surfaces with the orientation of the prismatic planes in the micropillar (Fig. 7f). Thus, the CRSSs for prismatic slip (obtained from the yield stress and the SF for both micropillar orientations) were determined to be 105 ± 4 MPa and 105 ± 2 MPa along [11$\bar{2}$0] and [10$\bar{1}$0] orientations, respectively.

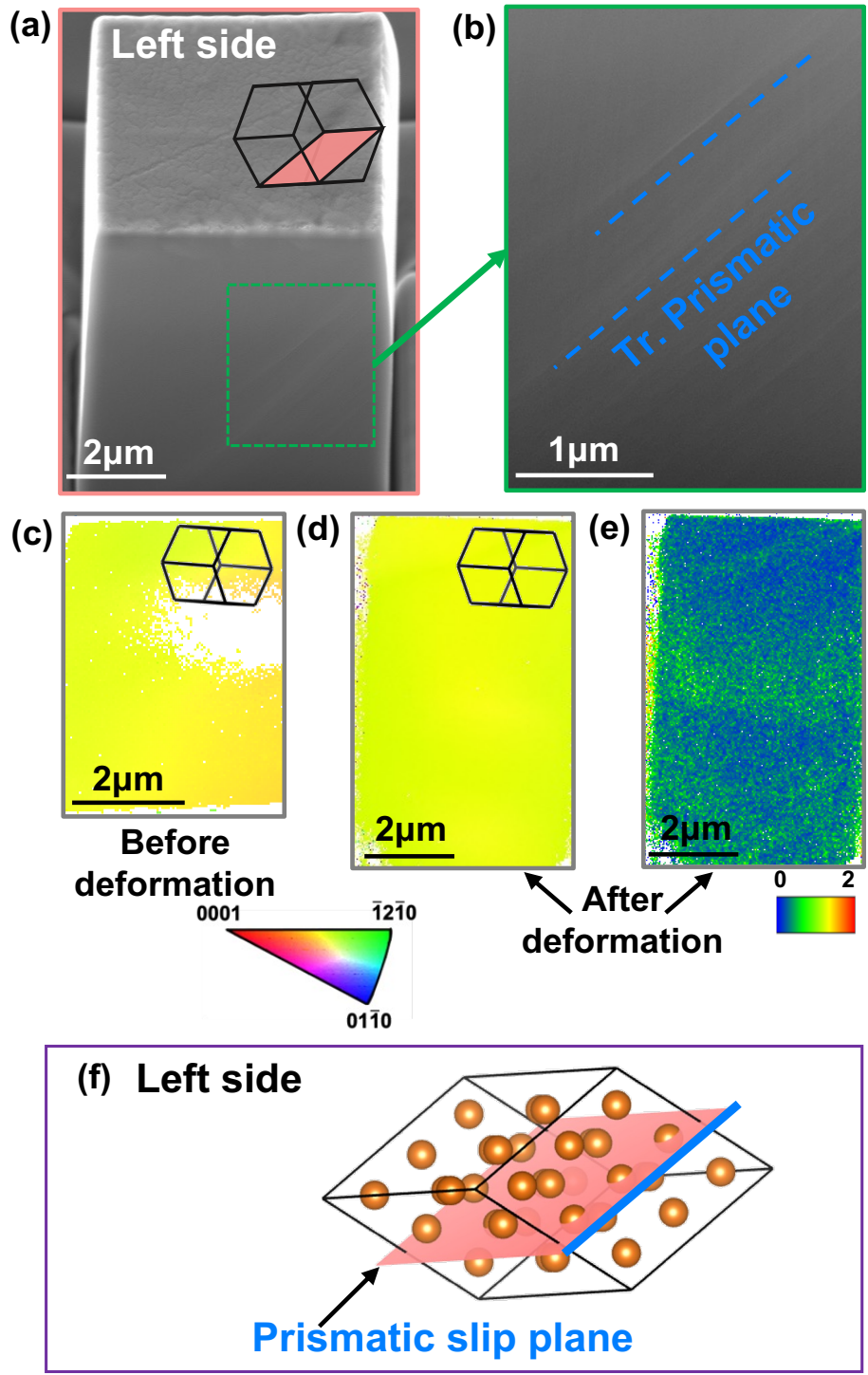

**Fig. 7.** SEM images of the micropillar deformed along [10$\bar{1}$0] from grain C. (a) Lateral left view side and (b) which shows the rectangular zone marked with a dashed line in Fig. 7a at higher magnification. The traces of the active slip planes are indicated by the blue dashed lines. (c) and (d) TKD maps of the lamella extracted from the undeformed region in the grain and along the compression direction from the deformed micropillar, respectively. (e) KAM map of the deformed micropillar in grain C. (f) Schematic of the crystallographic lattice showing the corresponding slip plane for the lateral left view in (a) and (b). The red plane indicates the theoretical prismatic glide plane, and the blue

line represents the corresponding traces.

Further assessment of the deformation mechanisms was carried out by means of TEM observations of the dislocation structures in a thin lamella extracted from the micropillar deformed along [10$\bar{1}$0] (Fig. 8). The lamella was nearly parallel to ($\bar{1}$2$\bar{1}$0) plane, as confirmed by the SADP in the inset in Fig. 8a, and there are no traces of twinning in the micropillar. Dark field micrographs of the square region marked in Fig. 8a are depicted in Figs. 8b and 8c with g = (10$\bar{1}$0) and g = (0002), respectively. Large density of dislocations is observed in Fig. 8b but they disappear from this region when g = (0002) in Fig. 8c. They are obviously <a> dislocations with 1/3 a [11$\bar{2}$0] or 1/3 [2$\bar{1}\bar{1}$0] Burgers vector, based on the dislocation extinguish condition. However, the SF of the {01$\bar{1}$0} [2$\bar{1}\bar{1}$0] prismatic slip system is very low (~0.05), thus, it is reasonable to assume that the Burgers vector of the <a> dislocations in Fig. 8b is 1/3 [11$\bar{2}$0]. The <a> screw dislocations are observed under g = (10$\bar{1}$0) condition as marked with yellow arrows in Fig. 8b. The Burgers vector of screw dislocation is parallel to the dislocation line, leading to the straight dislocation lines nearly parallel to trace of the basal planes (marked with a green line).

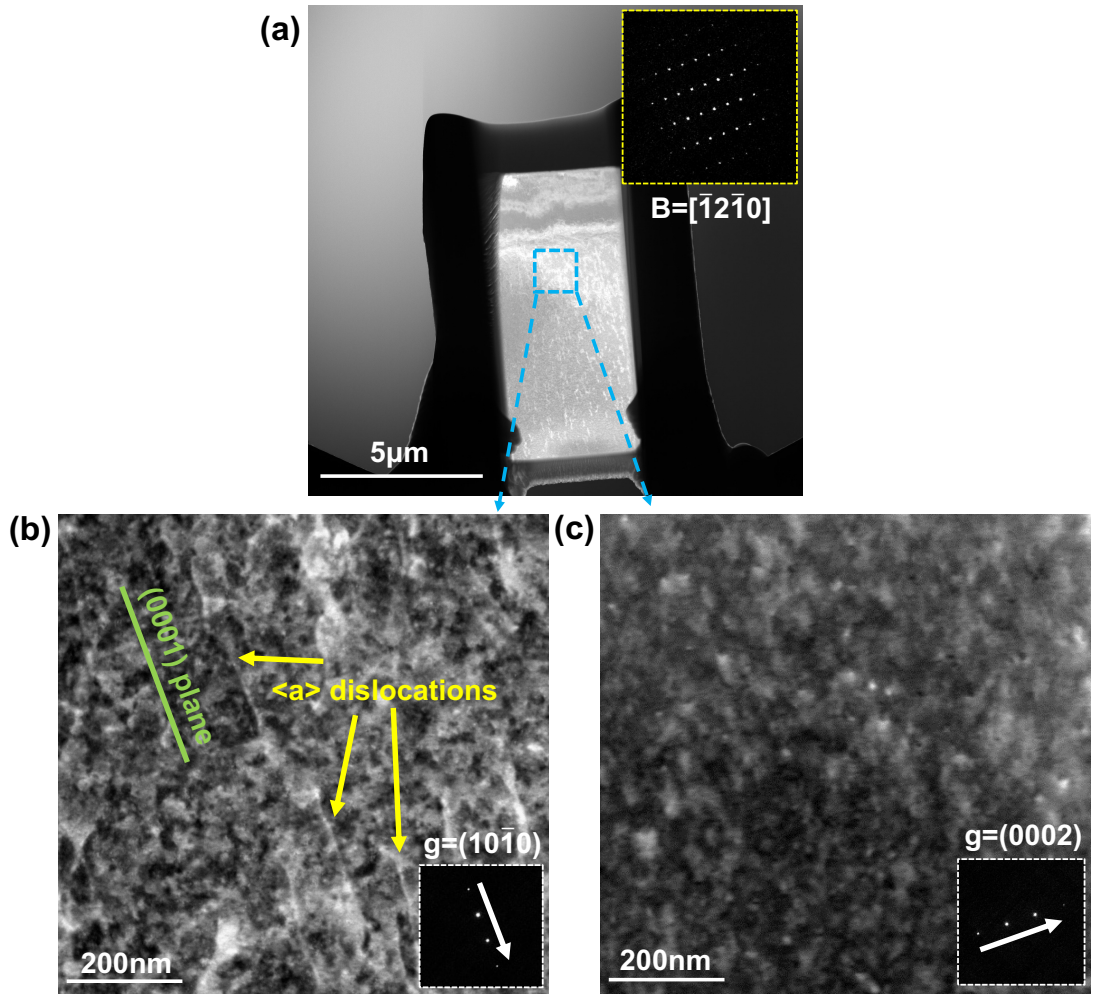

**Fig. 8.** TEM micrographs of the lamella extracted from the micropillar deformed along [10$\bar{1}$0]. The beam direction is parallel to [$\bar{1}$2$\bar{1}$0] orientation. (a) Low magnification view of the lamella. (b) and (c) High magnification dark field micrographs with g = (10$\bar{1}$0) and g = (0002), respectively, from the region marked with a blue square in (a).

The activation of the <a> prismatic slip during compression along the *a-axis* has been reported recently in Mg-Zn-Ca alloy (Wang et al., 2021b) in combination with tensile twinning. However, the activation of <a> prismatic slip and the suppression of tensile twinning during compression along the *a-axis* has not been found at ambient temperature in pure Mg (Y. Liu et al., 2017) or any Mg alloys (Li et al., 2021b, 2021a; Wang et al., 2021a, 2020; Wu et al., 2020). This result is very surprising because compression of Mg and its alloys along the *a-axis* (or equivalent extension along the *c-axis*) easily leads to the nucleation and growth of {10$\bar{1}$2} tensile twins, because the associated CRSS to promote tensile twin is much lower than that necessary to activate

<c+a> pyramidal slip or <a> prismatic slip. While the addition of Y and Ca in solid solution leads to a large increase in the CRSS for <a> prismatic slip with respect to pure Mg (from 39 MPa in pure Mg (Kaya, 2013) to 105 MPa), it seems to have a much larger effect on the CRSS for twin nucleation. In fact, considering the maximum stresses attained in the micropillar compression tests along $[11\bar{2}0]$ and $[10\bar{1}0]$ orientations (258 MPa in Fig. 5a and 303 MPa in Fig. 5b, respectively) and the maximum SFs for tensile twinning in both orientations (Table 1), it can be estimated that the CRSS for twin nucleation in the Mg-Y-Ca alloys should be higher than 148 MPa.

### 3.2.3 Deformation mechanisms in micropillar of grain D

The engineering stress-strain curves obtained from the compression micropillars carved from grain D along [0001] orientation are plotted in Fig. 9a. After the elastic region, a strong linear hardening was observed in the plastic region. The yield stress (marked by the purple stars in the inset) was 431 ± 15 MPa. This mechanical response is in good agreement with the results reported in Mg-0.4Y (wt. %) and Mg-4Y (wt. %) alloys (Wu et al., 2020) as well as in precipitation-hardened Mg-4Zn (wt. %) alloy (Alizadeh et al., 2021) under *c-axis* compression. In all these cases, the presence of Y in solid solution or of $\beta'_1$ precipitates increased the CRSS for basal slip and plastic deformation was accommodated through <c+a> pyramidal slip due to the low SF of basal planes in this orientation. The SEM micrograph of the lateral side of deformed micropillar in Fig. 9b, shows no slip traces but this behavior is also typical of pyramidal slip, which does not lead to visible slip traces on the micropillar surface. The orientation of the micropillar after deformation was assessed by TKD in a thin lamella extracted along the compression direction. The IPF map in Fig. 9c indicates the absence of the tensile twinning during deformation.

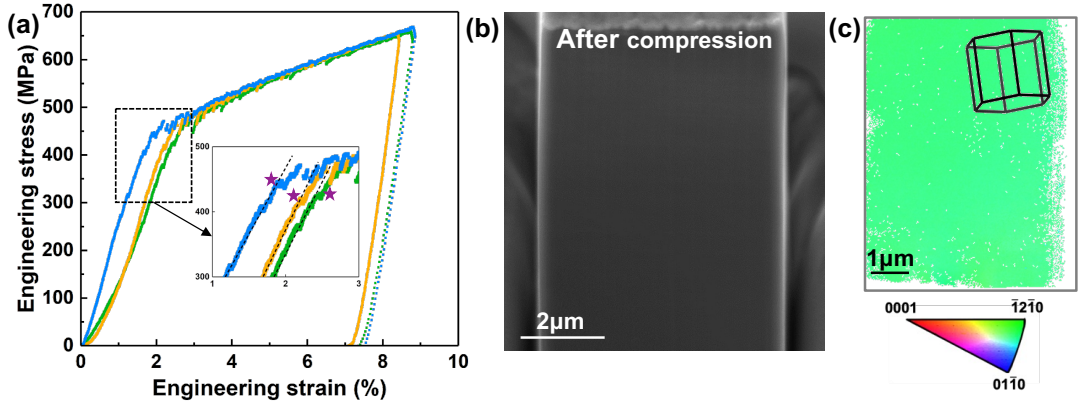

**Fig. 9.** (a) Engineering stress-strain curves from the micropillar deformed in compression along [0001] in grain D. (b) SEM micropillar of the lateral side of the deformed micropillar. (c) IPF map of the lamella extracted from the deformed micropillar.

To further elucidate the deformation mechanisms, the analysis of the dislocation structures was carried out by TEM in a thin lamella extracted from the deformed micropillar. The beam direction was parallel to [11$\bar{2}$0] as confirmed by the SADP in the inset in Fig. 10a. Two-beam condition imaging was performed with g = (0002) and g = (10$\bar{1}$0) and the corresponding dark field micrographs are depicted in Figs. 10b and Fig. 10c, respectively. A large density of <c+a> dislocations (marked with blue arrows) is observed under g = (0002) in Fig. 10b, and some <a> components are still in contrast at the same location when the operation vector changes to g = (10$\bar{1}$0) in Fig. 10c. The detail of the rectangular region marked with purple dashed lines in Fig. 10b is shown at higher magnification in Fig. 10d. The <c+a> dislocations (marked with the blue dashed lines) are [$\bar{1}\bar{1}$23]/3 and [1$\bar{2}\bar{1}$3]/3 according to the [11$\bar{2}$0] crystal orientation in Fig. 10e. These results are in agreement with those reported in Mg-Zn-Ca and Mg-Y alloys (Wang et al., 2021a; Wu et al., 2020). Nevertheless, it should be noticed that it is difficult to identify the active pyramidal plane, since both pyramidal I and pyramidal II planes contain the same slip directions. Thus, it can be concluded that plastic deformation along the *c-axis* in compression was dominated by <c+a> pyramidal dislocations. The activation of pyramidal slip was associated to homogeneous deformation and strong strain hardening (Basu et al., 2021). This high hardening rate is likely associated with short mean-free paths and this explains why no slip traces were found on the micropillar surface. It is not clear whether slip took place along pyramidal I plane or pyramidal II plane but the SF is slightly higher for pyramidal II for this particular orientation (Table 1), which is assumed to be the active one. Thus, the CRSS for <c+a> pyramidal II slip can be estimated as 203 ± 7 MPa from the SF and the yield stress.

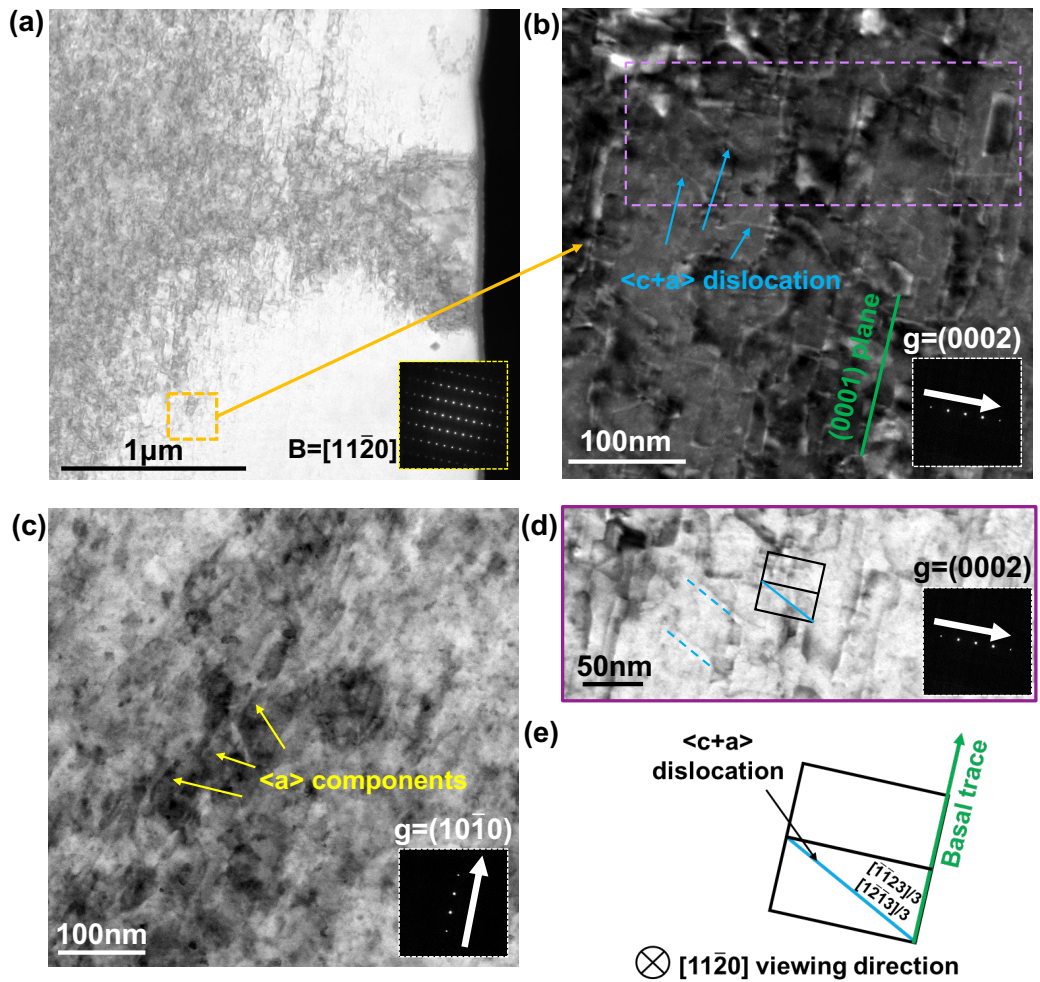

**Fig. 10.** (a) Bright field TEM image of the thin lamella extracted from the micropillar deformed along the [0001] orientation. The beam direction is [11$\bar{2}$0], as shown by the SADP in the inset. (b) and (c) Dark field TEM micrographs of the square region marked with orange dash lines in (a) under g = (0002) and g = (10$\bar{1}$0), respectively. (d) Bright field TEM micrograph obtained under g = (0002) from the rectangular region marked with purple dash lines in (b). The potential <c+a> pyramidal dislocations are indicated by the crystal orientation (black box) and the pyramidal slip trace (blue line). (e) Schematic of Mg crystal orientation and of the corresponding <c+a> dislocations (blue lines) from the [11$\bar{2}$0] projected view.

### 3.3 Effect of Y and Ca on the GSFE curves

The experimental evidence presented above shows that the addition of Y and Ca affects significantly the plastic deformation mechanisms. The changes in the deformation mechanisms are proposed to be associated with the modification of the slip resistance of the different slip systems due to the presence of the solute atoms. The GSFE is intimately associated with the activation barriers of the deformation modes, hence influencing their relative contributions to the overall deformation behavior (Sandlöbes et al., 2011). To ascertain the effect of the solute atoms (Y and/or Ca) on the slip activities, the GSFE ($\gamma$) curves were computed for the <a> slip systems in Mg-Y, Mg-Ca, and Mg-Y-Ca alloy, as well as in pure Mg for comparison.

The GSFE curves for {0001}<10$\bar{1}$0>, {1$\bar{1}$00}<11$\bar{2}$0> and {10$\bar{1}$1}<$\bar{1}$2$\bar{1}$0> slip systems are presented in Figs. 11a, 11b and 11c, respectively. The curves exhibited only one local maximum, from which the unstable stacking fault energy ($\gamma_{us}$) for each slip system was determined (Table 2). $\gamma_{us}$ is associated with the activation barrier for dislocation slip (Ding et al., 2018; Dong et al., 2018). Evidently, the addition of Y and Ca reduced slightly the $\gamma_{us}$ for <a> basal slip from 88 mJ/m$^2$ in pure Mg to a minimum of 64 mJ/m$^2$ in Mg-Ca or of 73 mJ/m$^2$ in Mg-Y and the $\gamma_{us}$ of Mg-Y-Ca (74 mJ/m$^2$) was similar with that of Mg-Y. However, the reduction in $\gamma_{us}$ for the <a> prismatic slip system was much more important, from ~235 mJ/m$^2$ in pure Mg to $\gamma_{us}$ of ~18 mJ/m$^2$ in the Mg-Y-Ca alloy (Table 2). This synergistic contribution of Y and Ca on $\gamma_{us}$ for <a> prismatic slip is obvious as the sole addition of either Y or Ca only reduced $\gamma_{us}$ to 120 mJ/m$^2$ (Table 2). The dramatic reduction of $\gamma_{us}$ for <a> prismatic slip in the Mg-Y-Ca alloy facilitates the activation of this deformation mechanism during plastic deformation. On the contrary, the $\gamma_{us}$ for <a> pyramidal I only changed from 304 mJ/m$^2$ in pure Mg to 318 mJ/m$^2$ in Mg-Y-Ca alloy. The sole addition of Ca (308 mJ/m$^2$) did not modify significantly $\gamma_{us}$ for <a> pyramidal I while Y (359 mJ/m$^2$) increased slightly $\gamma_{us}$ for <a> pyramidal I. Thus, <a> prismatic slip is favored by the addition of Y and Ca in comparison with <a> pyramidal I slip.

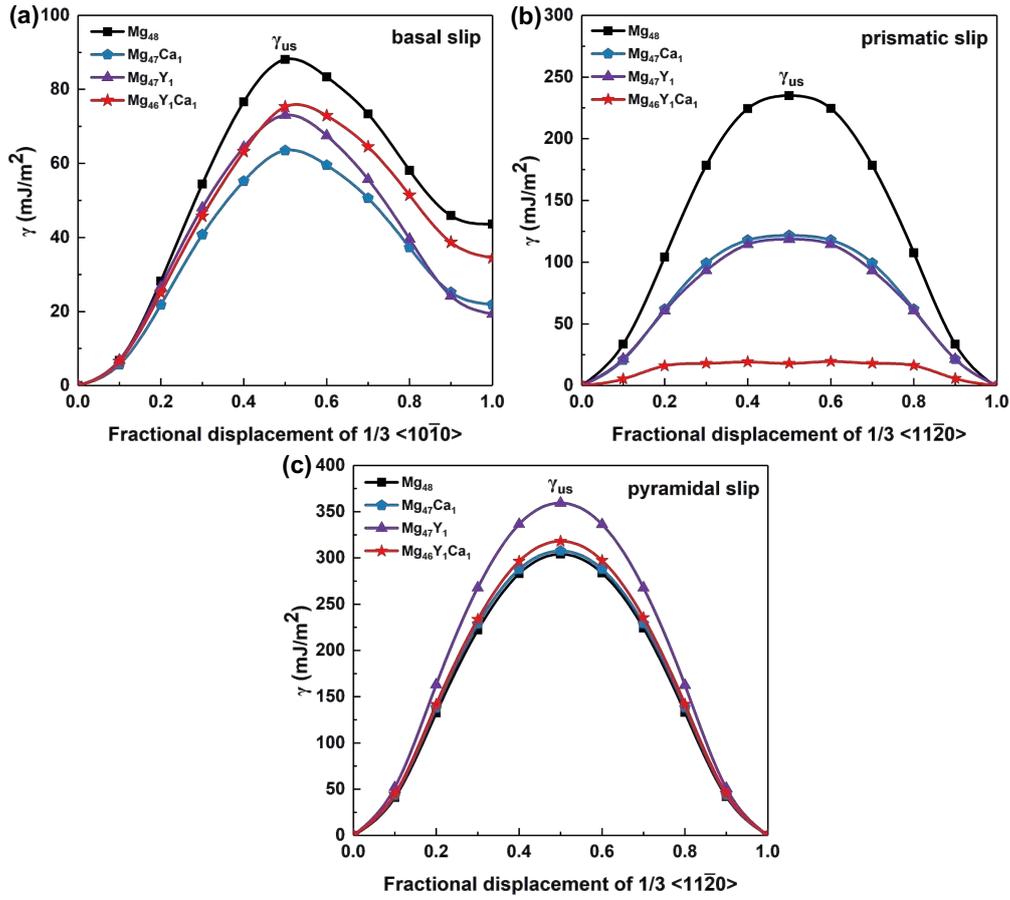

**Fig. 11.** Generalized stacking fault energy curves for (a) <a> basal (b) <a> prismatic, and (c) <a> pyramidal I slip systems in pure Mg, Mg-Ca, Mg-Y and Mg-Y-Ca alloys.

Table 2. The calculated $\gamma_{us}$, for basal <a>, prismatic <a>, and pyramidal I <a>, slip systems in the Mg-Ca, Mg-Y and Mg-Y-Ca alloys compared with pure Mg.

| Alloy | $\gamma_{us}$ (mJ/m$^2$) | | |
| --- | --- | --- | --- |
| | <a> Basal | <a> Prismatic | <a> Pyramidal I |
| $Mg_{48}$ | 88 | 235 | 304 |
| $Mg_{47}Ca_1$ | 64 | 121 | 308 |
| $Mg_{47}Y_1$ | 73 | 118 | 359 |
| $Mg_{46}Y_1Ca_1$ | 74 | 18 | 318 |

## 4. Discussion

### 4.1 Effect of Y and Ca on the CRSSs

The yield stresses measured from the micropillar compression tests in different orientations are summarized in Table 3. The CRSS for the dominant slip system in each orientation (following slip trace analysis and TEM characterization) is also presented in Table 3. They are <a> basal slip in the micropillars carved from grain A, <a> prismatic slip in the micropillars from grains B and C, and <c+a> pyramidal II slip in the micropillars from grain D. Moreover, twin nucleation was not observed in micropillars carved from grains B and C and this result can be used to obtain thresholds of the CRSS for twin nucleation from the maximum stress attained during the test and the maximum SF for tensile twinning in Table 1. These minimum values are also included in Table 3. It should be noticed that dimensions of the micropillars selected in this investigation follow previous results in Mg alloys (Li et al., 2021a; Wang et al., 2020; Wu et al., 2020) that indicate these values should not be very much influenced by the "smaller is stronger" effect reported for micropillar compression tests at the micron or sub-micron scale (Aitken et al., 2015; Chang et al., 2014).

Table 3. Yield stress and CRSS for different slip systems from micropillar compression tests along different orientation in the Mg-Y-Ca alloy.

| Grain | Loading direction | Yield stress (MPa) | CRSS (MPa) |
|---|---|---|---|
| A | [11$\bar{2}$3] | 65 ± 11 | 29 ± 5 (<a> basal slip) |
| B | [11$\bar{2}$0] | 219 ± 9 | 105 ± 4 (<a> prismatic slip)<br>> 111 MPa (tensile twin*) |
| C | [10$\bar{1}$0] | 228 ± 4 | 105 ± 2 (<a> prismatic slip)<br>> 148 MPa (tensile twin*) |
| D | [0001] | 431 ± 15 | 203 ± 7 (<c+a> pyramidal II slip) |

*: Tensile twin was not nucleated when the CRSS reached this value.

In order to ascertain the strengthening effect of Y and Ca atoms in solid solution, the CRSSs for the different slip systems in Mg-Y-Ca alloy are plotted in Fig. 12 along with those reported in the literature in pure Mg (Li et al., 2021a; Wang et al., 2019a), Mg-Al (Wang et al., 2020, 2019a), Mg-Zn (Li, 2019; Li et al., 2021a), Mg-Y (Li et al., 2021b; Wu et al., 2020), Mg-Zn-Ca (Wang et al., 2021a), Mg-Al-Ca (Luo et al., 2022) and Mg-Y-Zn (Chen et al., 2018) alloys. All these results were obtained from compression tests in micropillars with a cross-section around 5 × 5 μm² and, thus, size effects -if any- should not affect the comparison. The results for <a> basal slip in Fig. 12a show that the addition of Y in solid solution dramatically increases the CRSS in comparison with pure Mg (Li et al., 2021a; Wang et al., 2019a) and with Mg-Al (Wang et al., 2019a) or Mg-Zn (Li, 2019; Li et al., 2021a) alloys. These experimental data are supported by the first principles simulations of the solute/dislocation interaction energy, which showed the higher strengthening potential of Y for basal dislocations, in comparison with Al and Zn, because of the larger atomic radius and shear modulus misfit of Y with respect to Mg (Tehranchi et al., 2018). The addition of Ca to the Mg-Y does not increase the CRSS for basal slip according to our results while the strengthening effect of Ca in Mg-Al (Luo et al., 2022) or Mg-Zn (Wang et al., 2021a) is limited and may also be attributed to the elastic interaction between Ca solute atoms and dislocations.

Regarding <c+a> pyramidal slip (Fig. 12b), Zn and Al are the alloying elements which lead to the largest increase in the CRSS (Li et al., 2021a; Wang et al., 2020). Zn is more efficient but the solubility of Al in Mg is larger and CRSSs in the range of 200-250 MPa can be achieved for these binary alloys. Addition of 4 wt. % Y increases the CRSS up to 106 MPa (Wu et al., 2020) but the combination of Y and Ca leads to a CRSS of 203 ± 7 MPa, similar to the one found in the binary Mg-Zn alloy. Thus, Zn, Al and Y solutes increase the CRSS for <c+a> pyramidal slip due to the elastic interaction of the solutes with the dislocations, as in the case of <a> basal slip. It should be noticed that the activation and glide of <c+a> pyramidal dislocations is a complex process that also depends on dislocation dissociation during gliding due to the larger Burgers vector (Moitra et al., 2014; Tang and El-Awady, 2014). Atomistic simulations have shown that the presence of Y and Ca favors the activation for cross-slip/double cross-slip of <c+a> pyramidal dislocations, leading to new dislocation loops which can accommodate plastic deformation (Wu et al., 2018).

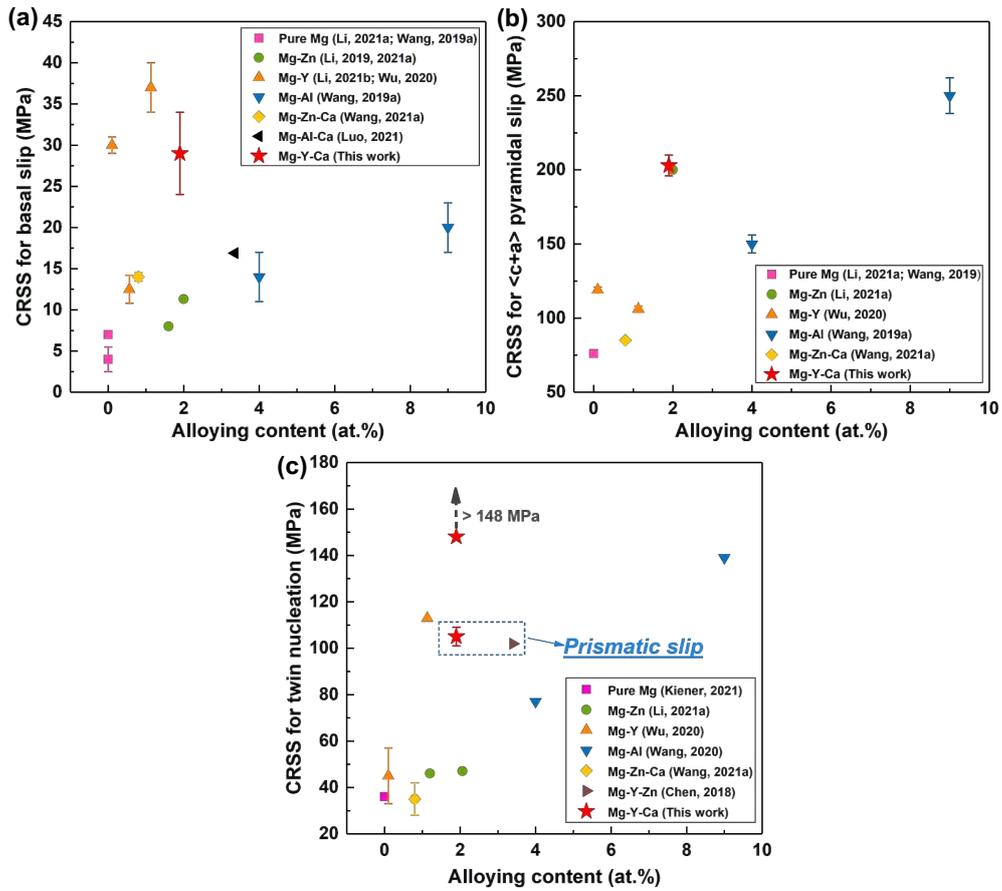

**Fig. 12**. CRSS for (a) basal slip, (b) pyramidal slip and (c) twin nucleation and prismatic slip in Mg and Mg alloys (Chen et al., 2018; Kiener et al., 2021; Li, 2019; Li et al., 2021b, 2021a; Luo et al., 2022; Wang et al., 2021a, 2020, 2019a; Wu et al., 2020), including the results obtained for Mg-Y-Ca alloy in this investigation. All data were obtained from compression tests in micropillars with a cross-section around 5 × 5 μm². The arrow in the CRSS for twin nucleation in Mg-Y-Ca indicates that the actual CRSS is higher than the value in the figure.

The CRSSs for tensile twin nucleation, measured by means of micropillar compression tests in Mg and different Mg alloys, are plotted in Fig. 12c (Kiener et al., 2021; Li et al., 2021a; Wang et al., 2020, 2021a; Wu et al., 2020). While the addition of Y (Wu et al., 2020) and Al (Wang et al., 2020) lead to the largest enhancements in the CRSS for twin nucleation (the latter because of the larger solid solubility), the highest CRSS is obtained for the ternary Mg-Y-Ca alloy which -following our experimental results- has to be higher than 148 MPa. Generally, the twin nucleation process is dominated by the dislocation-shearing and atomic shuffle. The strong strengthening provided by Y on the CRSS for twin nucleation can be ascribed to the inhibition of atomic shuffling due to the large atomic radius of Y (0.180 nm). Moreover, Ca has an even larger atomic radius (0.194 nm) and it is proposed that the synergistic contribution of both atoms in solid solution is responsible for the huge increase in the CRSS for twin nucleation. In addition, the elastic interaction of twinning dislocations with different solute atoms also leads to an increase in the CRSS for twin propagation (Ghazisaeidi et al., 2014; Stanford et al., 2015), as it has been reported in previous investigations (Li et al., 2021a, 2021b; Wang et al., 2020, 2021a). However, only the addition of Y and Ca can inhibit twin nucleation in micropillars suitable oriented for twinning, e.g., deformed in compression along [11$\bar{2}$0] and [10$\bar{1}$0] (Table 1).

The high CRSS for tensile twin nucleation in Mg-Y-Ca alloys leads to the activation the <a> prismatic slip, which becomes the dominant plastic deformation mechanism under *a-axis* compression. There is limited information on the CRSS for <a> prismatic slip (because either <a> basal slip or tensile twinning are usually activated before <a> prismatic slip to accommodate the plastic deformation) and the available experimental data on Mg-Y-Zn (Chen et al., 2018) (102 MPa) and Mg-Y-Ca (105 ± 4 MPa) are plotted in Fig. 12c. The CRSS for <a> prismatic slip is much lower than the CRSS for tensile twin nucleation in Mg-Y-Ca and, thus, tensile twinning is suppressed during compression parallel to the *a-axis*.

The CRSSs in Fig. 12 show that the strengthening effect of the Y and Ca for <a> prismatic slip is much lower than the ones reported for <c+a> pyramidal slip and twin nucleation, and also, in relative terms, for <a> basal. Moreover, evidence of <a> prismatic slip is unusual in Mg alloys except in the case of that they contain Ca (Zhu et al., 2019), indicating that the presence of Ca reduces the activation barriers for <a> prismatic slip glide. Besides, Chen et al., (2018) found that <a> prismatic slip was activated during micropillar compression testing of Mg-Y-Zn alloys but it was absent in solution-treated Mg-Zn alloys deformed along the same orientation (Li et al., 2021a; Wang et al., 2019b), implying that the addition of Y also facilitates prismatic slip. Although the elastic interaction of the solute atoms with <a> prismatic dislocations is expected to increase the CRSS, the reduction of the stacking fault energy due to the presence of Y and Ca reduces the activation barrier for dislocation movement on the slip plane and facilitates the activation of this slip system.

Overall, the addition of Y and Ca leads to a marked solid solution strengthening for <a> basal and <c+a> pyramidal slip as well as for the nucleation of tensile twins but not for <a> prismatic slip.

**4.2 Effect of plastic anisotropy on the ductility**

In general, the tensile ductility and formability of Mg alloys during the plastic deformation is dictated by the CRSS ratio between different slip systems, especially between non-basal and basal slip, the latter being the dominant deformation mechanism in most cases (G. Liu et al., 2017; Zhu et al., 2019). Therefore, the tensile ductility of different Mg alloys is plotted as a function of the CRSS ratios between different slip systems in Fig. 13 (Habibi et al., 2012; Huang et al., 2018; Shi et al., 2020; Wang et al., 2021b; Wu et al., 2010; Yang et al., 2022; Zhao et al., 2019a, 2019b; Zhu et al., 2020, 2019). The CRSS ratios were measured via micropillar compression tests in most of the alloys (Agnew et al., 2003; Li et al., 2021a; Shang et al., 2021; Wang et al., 2021a, 2020, 2019a, 2018; Wu et al., 2020; Zhu et al., 2019) with a few exceptions. The CRSS ratio between <a> prismatic and <a> basal slip in Mg-5Y (wt. %) (Huang et al., 2018) and Mg-0.47Ca (wt. %) (Zhu et al., 2019) were obtained from mechanical tests in polycrystals via slip trace analysis. Besides, those for pure Mg (Agnew et al., 2003), Mg-0.5Ca (wt. %) (Shang et al., 2021) and Mg-3Y (wt. %) (Wang et al., 2018) alloys were determined by the elasto-plastic self-consistent model, crystal plasticity finite element simulations, and the elastic viscoplastic self-consistent model, respectively. Moreover, the tensile elongation data were collected from pure Mg and wrought Mg alloys with similar grain sizes.

In general, reduced ratios between the CRSS for non-basal slip and basal slip are strongly associated with the improvement of the ductility of Mg alloys. This trend agrees with the data plotted in Fig. 13b, which shows a clear link between the reduction of CRSS $_{<c+a>\,pyramidal}$ / CRSS $_{<a>\,basal}$ and the increase in tensile elongation. However, the limited data of the influence of CRSS $_{<a>\,prismatic}$ / CRSS $_{<a>\,basal}$ on the tensile ductility in Fig. 13a are not conclusive. Obviously, low CRSS $_{<c+a>\ pyramidal}$ / CRSS $_{<a>\ basal}$ ratios favor isotropic deformation and limit the development of strong basal textures and both processes help to improve ductility and formability because activation of <c+a> dislocations benefits the strain accommodation along the *c-axis* (Liu et al., 2019). Besides, Wu et al., (2018) predicted that the addition of Y/Ca could significantly reduce the cross-slip energy barriers between pyramidal I and pyramidal II planes, thus promoting <c+a> dislocation cross-slip. Enhanced non-basal slip activities and cross-slip induce homogeneous deformation and improve the ductility.

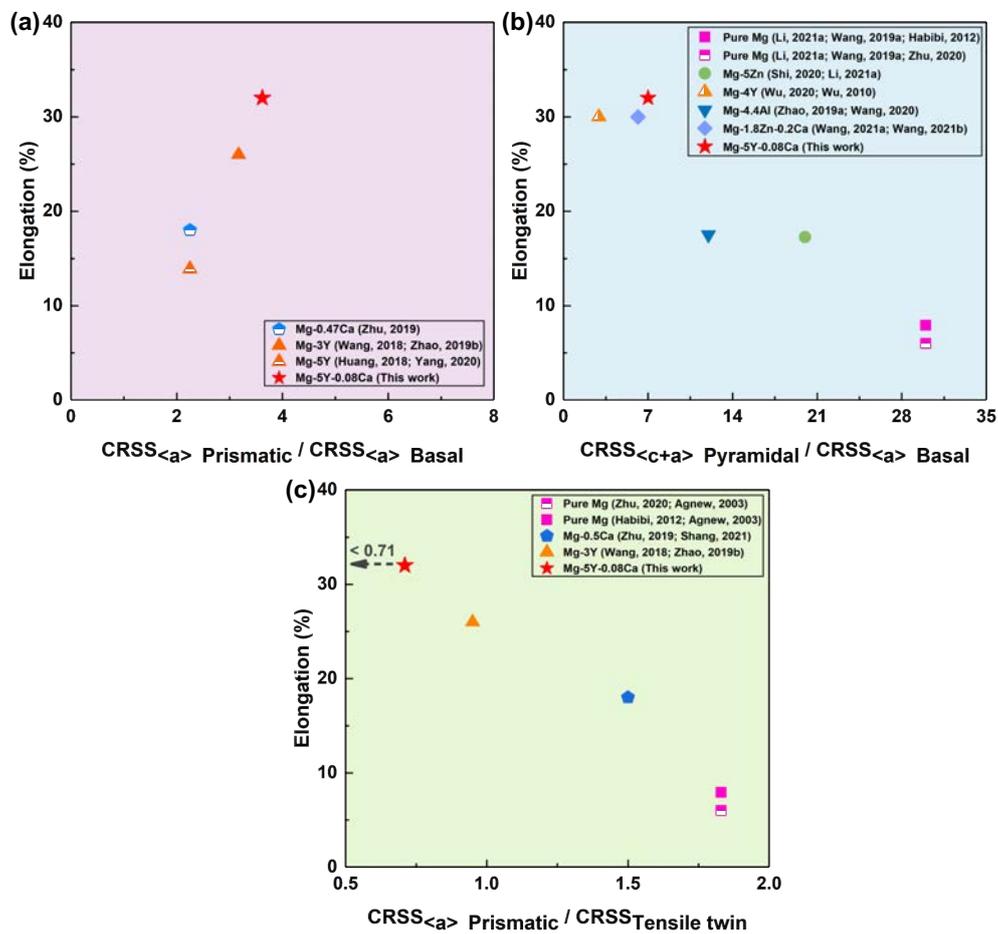

**Fig. 13.** Relation between the CRSS ratios of different slip systems (Agnew et al., 2003; Li et al., 2021a; Shang et al., 2021; Wang et al., 2021a, 2020, 2019a, 2018; Wu et al., 2020; Zhu et al., 2019) and the tensile elongation (Habibi et al., 2012; Shi et al., 2020; Wang et al., 2021b; Wu et al., 2010; Yang et al., 2022; Zhao et al., 2019a, 2019b; Zhu et al., 2020, 2019) in pure Mg and Mg alloys: (a) CRSS $_{<a>\ prismatic}$ / CRSS $_{<a>\ basal}$, (b) CRSS $_{<c+a>\ pyramidal}$ / CRSS $_{<a>\ basal}$, (c) CRSS $_{<a>\ prismatic}$ / CRSS $_{tensile\ twin}$.

It should be noted that these mechanisms are particularly relevant in Mg-Y (Wang et al., 2018) and Mg-Zn-Ca (Wang et al., 2021a, 2021b) alloys as well as in the Mg-Y-Ca alloy analyzed in this investigation. In all these cases, the presence of Y and/or Ca also leads to a high increase in the CRSS for twin nucleation while the CRSS for <a> prismatic slip is not strongly affected. As a result, the CRSS $_{<a>\text{ prismatic}}$ / CRSS $_{\text{tensile twin}}$ is dramatically reduced and this is accompanied by a large increase in the tensile ductility, as shown in Fig. 13c. Particularly, tensile twinning is replaced by <a> prismatic slip during compressive deformation along the *a*-axis if CRSS $_{<a>\text{ prismatic}}$ / CRSS $_{\text{tensile twin}}$ < 1 and twinning only occurs in grains deformed in tension along the *c*-axis. Moreover, as the CRSS for <a> prismatic slip is smaller than that for <c+a> pyramidal slip, the former becomes the dominant plastic deformation mechanism in grains suitable oriented for both. It should be noted that <c+a> pyramidal slip is associated with a large strain hardening (Fig. 9a) that it is not present for <a> prismatic slip (Fig. 5). Thus, pyramidal slip induced large stress concentrations at grain boundaries that facilitate the nucleation of damage but this process is not activated if <a> prismatic slip is dominant.

In general, the preferential activation of basal slip and tensile twinning during processing always introduces a strong basal texture in wrought Mg and Mg alloys, leading to the plastic anisotropy, crack formation and limited ductility (Sabat et al., 2015; Wang et al., 2021a). The addition of Y and Ca in our alloy strongly enhanced the activation of prismatic <a> and pyramidal <c+a> slip, which also contribute to reduce the intensity of the texure during extrusion, as shown in Figs. 2a and 2b. This limited texture also contributes to reduce the plastic anisotropy.

It should also be noted that the overall mechanical response of polycrystals cannot fully ascertained by means of micromechanical tests in single crystals because other factors (grain boundaries, grain size and texture) play a key role in the mechanical response. However, it should be emphasized that the plastic deformation of each crystal within the polycrystal is intrinsically related to that of a single crystal (Wang et al., 2021a) and, hence, it is important to ascertain the plastic deformation mechanisms in single crystals to understand the complex mechanisms in bulk polycrystalline samples.

Overall, these results indicate that the presence of Y and Ca in solid solution in Mg alloys leads to a large increase in the CRSS for <a> basal slip (which induces a large reduction in CRSS $_{<c+a>\,pyramidal}$ / CRSS $_{<a>\,basal}$) while CRSS $_{<a>\,prismatic}$ / CRSS $_{tensile\,twin}$ < 1. As a result, plastic deformation in polycrystals in more isotropic and localization of the deformation in the form intense basal slips that promote fracture is suppressed (Sandlöbes et al., 2011). Moreover, twinning and <c+a> pyramidal slip are replaced by <a> prismatic slip in grains deformed along the *a-axis*. Suppression of twinning (which induces strong anisotropy in the plastic deformation in textured alloys) and the activation of <a> prismatic slip (which provides an additional plastic deformation mechanism with limited hardening) lead to an important improvement in the tensile ductility of Mg alloys.

## 5. Conclusions

The deformation mechanisms of a Mg-5Y-0.08Ca (wt. %) alloy, with a superior tensile elongation (32%), were studied by means of micropillar compression tests, slip trace analysis along different orientations, TEM as well as TKD. It was found that the presence of Y and Ca in solid solution led to a huge increase in the CRSS for <a> basal slip (29 ± 5 MPa), <c+a> pyramidal slip (203 ± 7 MPa) and tensile twin nucleation (above 148 MPa). This behavior was attributed to the large mismatch of the atomic radii and elastic constants of the Y and Ca atoms with respect to Mg, which leads to a strong interaction of the dislocations with the solute atoms and hinders atomic shuffling, that is necessary to activate twin nucleation. On the contrary, the CRSS for <a> prismatic slip only increases up to 105 ± 4 MPa because the hardening induced by the interaction of the solute atoms with dislocations is partially balanced by the reduction in the stacking fault energy associated with <a> prismatic slip due to the presence of Y and Ca.

The changes in the CRSS for slip and tensile twinning in Mg-Y-Ca alloys modify the dominant deformation mechanisms. In particular, the CRSS $_{<a>\,prismatic}$ / CRSS $_{tensile\,twin}$ is dramatically reduced and tensile twinning is replaced by <a> prismatic slip during compressive deformation along the *a-axis* if CRSS $_{<a>\,prismatic}$ / CRSS $_{tensile\,twin}$ < 1. Moreover, as the CRSS for <a> prismatic slip is smaller than that for <c+a> pyramidal slip, the former becomes the dominant plastic deformation mechanism in grains suitable oriented for both. As a result, reduction of twinning (which induces strong anisotropy in the plastic deformation in textured alloys) and the activation of <a> prismatic slip (which provides an additional plastic deformation mechanism with limited hardening) lead to an important improvement in the tensile ductility of Mg alloys.

**Acknowledgements**

This work was supported by the National Natural Science Foundation of China (Grant Nos. 52001199 and 51825101). Y. Cui acknowledges the support from the Shanghai Sailing Program (Grant No. 22YF1419300). JLL acknowledges the support from the Spanish Ministry of Science (HexaGB project, reference RTI2018-098245) and from the MAT4.0-CM project funded by the Comunidad de Madrid under programme S2018/NMT-4381.